# The Quantum-Classical Transition and Wave Packet Dispersion

## C. L. Herzenberg

**Abstract**
*Two recent studies have presented new information relevant to the transition from quantum behavior to classical behavior, and related this to parameters characterizing the universe as a whole. The present study based on a separate approach has developed similar results that appear to substantiate aspects of earlier work and also to introduce further new ideas.*

## 1. INTRODUCTION

The question of why our everyday world behaves in a classical rather than quantum manner has been of concern for many years. Generally, the assumption is that our world is not essentially classical, but rather quantum mechanical at a fundamental level. Various types of effects that may lead to classicity or classicality in quantum mechanical systems have been examined, including decoherence effects, recently in association with coarse-graining and the presence of fluctuations in experimental apparatus.[1,2] More recently, possible close connections of the transition from quantum to classical behavior with certain characteristics of the universe as a whole have been investigated.[3-6] In the present paper, we will examine such effects further in the context of the dispersion of quantum wave packets.

## 2. HOW DOES QUANTUM BEHAVIOR TURN INTO CLASSICAL BEHAVIOR?

First we will revisit briefly the question of how quantum mechanical behavior can turn into classical mechanical behavior. In classical mechanics, objects are characterized by well-defined positions and momenta that are predictable from precise initial data. In the classical limit of quantum mechanics, we would therefore anticipate the presence of a compact probability distribution corresponding to a compact wave packet.[7] However, a compact probability distribution cannot be achieved with a single eigenfunction, because the spatial part of an eigenfunction is never compact; it always fills the classically available space, that is, the region in which the potential energy is less than the energy eigenvalue.[7] Furthermore, the expectation values for a single eigenfunction are time-independent; they describe a stationary state. Thus, the classical limit necessarily involves superposition.

Thus, for classical behavior, the region in which an object could be or is most or quite likely to be present must be compact, and for the case of an extended object, it would seem that it should not extend beyond the object's physical extent. Thus, our approach is based on the criterion that when the probability distribution describing the object extends considerably beyond the object's size or physical extent, then the object can be expected to behave quantum mechanically; whereas, if the probability distribution associated with the object is largely confined within its boundaries, the object will behave in a predominantly classical manner.

## 3. SUMMARY OF RECENT STUDIES

These recent studies appear to demonstrate effects on quantum behavior dependent on the age or expansion rate of the universe. These studies have led to the conclusion that quantum mechanical objects must, if sufficiently large, be described by classical mechanics. Thus, above a threshold size, objects may be regarded as undergoing a quantum-classical transition.[3-6]

These studies have also provided straightforward interpretations of some possible aspects of the quantum-classical transition. In the first of these investigations, calculations suggested that the Hubble flow due to the expansion of space within any extended object would introduce velocity and momentum uncertainties which would be coupled to a size-dependent uncertainty in position that is large for small objects and small for large objects, thus identifying an intermediate size that distinguishes quantum behavior from classical behavior. In a subsequent separate investigation, calculations suggested that in stochastic quantum mechanics, the extent of diffusion of a quantum object over the lifetime of the universe would define limits on the spatial extent of quantum behavior for a physical object. In both cases, the critical or threshold size was shown to be similarly dependent on an object's mass and on the Hubble constant.[3,6]

The present paper is a supplementary study that is based on a somewhat different approach than the previous work and yet leads to a similar result, based on relevant aspects of the behavior of compact wave packets.

For orientation and comparison purposes and to examine possible issues, we will look briefly at each of the earlier studies.

## 4. DISCUSSION OF HUBBLE FLOW APPROACH

In a preceding study, the effect of the Hubble flow in introducing velocity and momentum uncertainties within the space occupied by an extended object was examined.[3] The spread of velocities associated with Hubble expansion at the various different locations within an extended object can be estimated. Since the cosmic expansion velocity between two spatial positions is proportional to the distance separating them, the proportionality factor being the Hubble constant, the full spread in velocities along any direction within an object would be expected to reach an approximate value:

$$\Delta v \approx H_o L \qquad (1)$$

Here, $H_o$ is the Hubble constant, L is a length characterizing the size of the object, and $\Delta v$ is the spread in Hubble expansion velocities within the object. This spread in velocities at different locations in the object was treated as providing an estimate of an uncertainty in velocity associated with the object. An estimate of the uncertainty in momentum $\Delta p$ associated with this uncertainty in velocity could be calculated by forming the product of the mass m of the object and the uncertainty in velocity associated with the object as a whole; then, by invoking the Heisenberg uncertainty principle in the form $\Delta p \Delta x \geq h/4\pi$, where h is Planck's constant, an approximate limit on an associated uncertainty in position was evaluated as:

$$\Delta x \geq h/(4\pi m H_o L) \qquad (2)$$

Here, m is the mass of the object. When the uncertainty in position is at its minimum value corresponding to the equals sign in Eqn. (2), the positional uncertainty has a minimum value allowable under the uncertainty principle in the presence of a Hubble spread of velocities within the object, within the approximations of the derivation.

As noted earlier, if the uncertainty in position exceeds the physical size of the object, the object may be expected to be governed by quantum mechanics, whereas if the uncertainty in position is smaller than the physical size of the object, the object would be expected to behave largely classically. A threshold size or critical length was then introduced as the length at which the minimum uncertainty in location is just equal to the length associated with the object. This led to identifying as a critical length the quantity:

$$L_{cr} = [h/(4\pi m H_o)]^{1/2} \qquad (3)$$

In this approach, a threshold or critical size is determined by the Hubble velocities present in the space occupied by the extended object, but is expressed in terms of the object's mass and the Hubble constant.

While the original derivation of such a threshold or critical size for a quantum-classical transition attracted some interest, informal criticism has been leveled at its attendant derivation for several reasons. Initially, some concern existed as to whether it would be appropriate to treat the spread in Hubble velocities within an object as an uncertainty in velocity leading to an uncertainty in momentum; however this concern seems to have abated.[5] In addition, concern was expressed because of the fact that, while Hubble flow is manifestly present on a cosmological scale, the presence of local Hubble flow remains disputed. A division of opinion exists on this subject, with some members of the scientific community regarding local systems as participating in the Hubble flow, with others taking the position that local systems would not participate, or would not participate fully, in the Hubble flow.[3]

It may be noted that if partial suppression of local Hubble expansion were to occur, this would lead to larger estimates for the minimum uncertainty in position; and Eqn. (3) for the critical length at which the quantum-classical transition would take place would not be accurate. The conclusion would still be valid that sufficiently large objects would have to behave classically; however, the critical size would be larger than the value derived for the presence of full scale Hubble flow. This situation could be treated formally in terms of a modified local Hubble-type expansion coefficient corresponding to a slower expansion or longer Hubble time than is actually observed for the universe as a whole.

The issue of possible local suppression of Hubble flow is also somewhat puzzling in this context because it would tend to suggest that, for example, a free object that is gravitationally unbound and located in outer space remote from other objects (and hence presumably participating in full Hubble expansion) would undergo a quantum-classical transition differently than would an otherwise similar object located in a gravitationally bound system (where it might for that reason experience local suppression of Hubble expansion).

In an effort to clarify the situation, the question of possible cosmic effects on the quantum-classical transition was also examined from another point of view.

## 5. DISCUSSION OF STOCHASTIC QUANTUM THEORY APPROACH

Stochastic quantum theory was examined in this context in order to provide a different approach to the question of the quantum-classical transition.[4] Calculations based on stochastic quantum theory showed that the diffusion of a quantum object as a function of time determines a size domain within which the object exhibits stochastic quantum behavior and outside of which that is essentially absent. It was shown that over the lifetime of the universe, a quantum particle or object would undergo stochastic diffusion over a spatial region of size approximately equal to the critical length associated with the quantum-classical transition. Thus, the critical size obtained by this method is essentially the same as that obtained by the earlier approach based on the presence of Hubble velocities.[4]

Stochastic interpretations of quantum mechanics generally treat quantum mechanics as fundamentally a classical theory of inherently probabilistic or stochastic processes, with the essentially quantum features arising from the stochasticity. In stochastic quantum mechanics it is found that linearized equations formally identical with the Schroedinger equation can be obtained if the diffusion coefficient is taken as equal to $h/4\pi m$.[4] In the case of stochastic diffusion of a quantum object in three dimensions, the root mean square distance or square root of the variance would be given by:

$$x_d(t) = (3ht/2\pi m)^{1/2} \qquad (4)$$

This equation provides us with an evaluation of the distance through which a quantum object may be expected to diffuse during an interval of time t. The root mean square distance reached in the random walk or diffusion of a quantum object would thus depend on the square root of the time available. If this time is taken to be the Hubble time, that is, the inverse of the Hubble constant, then the root mean square diffusion distance for a quantum object over the Hubble time is found to be:[4]

$$x_d(t_{Ho}) = (3h/2\pi mH_o)^{1/2} \qquad (5)$$

These results appear to tell us that during the time that has elapsed since the Big Bang, a quantum particle or object would undergo stochastic diffusion over a spatial region having a size comparable to the distance specified in Eqn. (5).

We again relied on the criterion that if this intrinsically quantum, non-classical region is larger than the physical size of the object, the object may be expected to behave quantum mechanically; while if this region is smaller than the physical size of the extended object, the object may be expected to behave classically.

Comparison of the diffusion distance for a quantum object in stochastic quantum mechanics with the physical size of the object then led to a substantially similar result to that obtained from Hubble flow.[4] The estimate of the critical length estimate $L_{cr}$ associated with the quantum-classical transition based on the uncertainty principle and the effect of Hubble velocities in an extended object, and the estimate made for diffusion displacement over the Hubble time turned out to be quite similar quantities with the same functional dependence on all of the same parameters, differing in value by only small numerical factors. However, the stochastic interpretation of quantum theory is not universally accepted, so these results based on stochastic quantum theory were also not fully convincing to all members of the research community.

While this approach provides a result in essential agreement with that obtained from the approach based in Hubble velocities, it also it appears to exhibit some additional interesting implications. Eqn. (5) is based on a time comparable to the age of the universe, which presumably would be applicable to stable particles and other objects created early in the history of the universe. However it seems possible that other values of the time might be applicable to other types of object created more recently. Thus, objects assembled more recently might require shorter times, and hence be characterized by smaller regions exhibiting quantum behavior. Such more recently assembled objects would accordingly be characterized by smaller effective critical lengths. It would appear that recently formed objects might thus be expected to behave more classically, while longer-lived objects might be expected to behave more quantum mechanically, in accordance with Eqns. (4) and (5). This would appear to suggest, for example, that an elementary particle recently emerged from a collision or disintegration might behave in a somewhat more classical manner that an otherwise identical particle that has been present from the beginning of the universe. This somewhat startling idea might give one pause, and it would seem that perhaps some exemptions or modifications might be required.

However, consideration of the behavior of initially confined wave packets may clarify this possible concern to some extent.

## 6. A THIRD APPROACH: QUANTUM WAVE PACKET BEHAVIOR

As a supplement to previous approaches and in order to provide additional evidence for and insight into these potentially fundamental results on the quantum-classical transition, an additional argument is now presented, based explicitly on wave packet behavior in ordinary quantum mechanics.

## 7. THE CASE OF A GAUSSIAN WAVE PACKET: AN ESTIMATE BASED ON WAVE PACKET EVOLUTION

We can also address the quantum-classical transition in the context of the probability distribution associated with a dispersing wave packet by examining the time evolution of a free wave packet describing an object not subject to external forces.

Since for a Gaussian wave packet, the calculations can be carried out exactly, we will work with Gaussian wave packets.

To simplify the analysis, we will consider a one-dimensional wave packet with a Gaussian distribution of wave numbers. We will start with an initial wave packet, and examine its dispersion to find out what happens at later times.

An initial Gaussian wave packet would have the form:[8]

$$\Psi(x,0) = A \exp[-x^2/2\sigma^2(0)] \, e^{ikx} \qquad (6)$$

Here, $\sigma(0)$ is the standard deviation with respect to x at an initial time zero, k is the mean wave number, and A is a normalization coefficient.

As time goes on, the quantum wave packet will disperse, since each plane-wave component of the wave packet has a different wave number and therefore will propagate at a different velocity. Thus, all of the plane wave components move away at different rates and the probability distribution associated with the wave packet becomes more spread out or dispersed with time.

The width of the wave packet will increase with time from its initial value, and the time dependence can be calculated and can be shown to be:[8]

$$\sigma(t) = [\sigma^2(0) + \{ht/4\pi m\sigma(0)\}^2]^{1/2} \qquad (7)$$

The subsequent width of the wave packet can be seen to depend on the value of the time, the mass of the object, and in this case of a Gaussian wave packet, the initial wave packet width. As the spatial extent of the wave packet increases, the probability of finding the particle within an infinitesimal interval of its mean position steadily decreases with time as the wave packet disperses.

Let us examine the characteristics of a Gaussian wave packet at a time T after its inception. We can inquire what the minimum width of a Gaussian wave packet at a particular time T will be. For a particular value of the time, T, the wave packet width can be seen to be a minimum for the case of a wave packet having an initial width:

$$\sigma(0) = (hT/4\pi m)^{1/2} \qquad (8)$$

If we insert the minimization criterion in Eqn. (8) into Eqn. (7), we find that the minimum width that a Gaussian wave packet could have at a time T will be given by:

$$\sigma(T) = (hT/2\pi m)^{1/2} \qquad (9)$$

While this result has been obtained for Gaussian wave packets, it can provide an estimate at least of the spatial extent of a compact wave packet probability distribution more generally and hence provide an estimate of the spatial extent of the quantum behavior associated with a quantum object.

What does this result tell us about a quantum wave packet that has been spreading out over cosmic time? We can estimate approximately the spatial extent of a quantum object that has had the lifetime of the universe during which to disperse by inserting the Hubble time $T_H = 1/H_o$ into Eqn. (9). This leads to an estimate for the present value of the wave packet width as:

$$\sigma(T_H) = (h/2\pi m H_o)^{1/2} \qquad (10)$$

So we find that over the lifetime of the universe, a quantum object characterized by a Gaussian wave packet of minimum size would be dispersedover a spatial region of size approximately equal to previously derived values for the critical length associated with the quantum-classical transition.

As noted earlier in connection with results from diffusion behavior in stochastic quantum theory, we should also think about the possible significance of the results for times shorter than the age of the universe. Eqn. (10) is based on a wave packet dispersion time comparable to the age of the universe, which presumably would be applicable to stable particles created early in the history of the universe, however again it seems possible that other times might apply to more recently formed objects. Objects assembled more recently might be more properly described by shorter times, and hence be described by smaller wave packet widths that would characterize smaller regions exhibiting quantum behavior. Thus it would appear that recently formed objects might be expected to behave in a more classical manner, while longer-lived objects might be expected to behave more quantum mechanically, in accordance with Eqns. (9) and (10).

## 8. FURTHER DISCUSSION AND CONCLUSIONS

In each of the three recent approaches used for examining the quantum-classical transition, we compare the physical size of the extended object with the size of the region in which its quantum behavior appears to be expressed. The reason that the results from these three different approaches are largely in agreement arises from the fact that the region in which quantum behavior is expressed turns out to be approximately the same in all three approaches.

In the static case of ordinary quantum mechanics, examined using the Heisenberg uncertainty principle, the uncertainty in spatial location (which describes the region within which quantum behavior takes place) depends on the momentum spread associated with the object. In this case, the momentum spread is present because of the fact that the object is extended and hence incorporates a spread of Hubble velocities. In this application of conventional quantum mechanics using the uncertainty principle approach, we are dealing with what amounts to a steady-state situation. The object exists; the presence of the Hubble spread of velocities within it is affecting the extent of its quantum behavior, but we do not inquire into its history.

In the stochastic quantum theory approach, the region associated with quantum behavior in which the particle is found is the region in which diffusion has occurred. In the application of stochastic interpretations of quantum mechanics, we consider that the particle has been in stochastic motion throughout its history. The region in which its stochastic behavior is expressed will be limited to the region in which it has diffused throughout its history, and the object's quantum behavior is then limited to the region in which its stochastic behavior is expressed. Here the diffusion behavior is explicitly time-dependent, and we visualize it as having been occurring since the beginning of the universe at the Big Bang with the current status as the result of that history. In the diffusion approach, it would seem that the size of the region encompassing quantum behavior is somehow cumulative over history. What size it is now depends on the stochastic motion, and being an extended object is not requisite, and expansion of the universe is not explicitly required.

In the wave packet approach, the region associated with quantum behavior is the present size of the wave packet that has evolved and dispersed over time. Without the requirement of localization, the object would presumably behave as a pure monochromatic wave extending throughout all space. However, if an initial localization has been applied to form the wave packet, it will subsequently disperse, forming a wider wave packet at the present time. Explicit expansion of the universe is not required in this approach either.

As noted earlier, these results introduce an additional question in regard to what dispersion time may be relevant: while roughly the age of the universe may be expected to apply to many basic objects such as nuclei and atoms formed relatively long ago, a smaller time interval may apply for objects formed in the more recent past, and that would lead to a smaller wave packet width. This would suggest the possibility that effective critical lengths might be dependent not only on an object's mass and the Hubble constant, but might also exhibit some dependence on the lifetime of the object itself. But in all cases, sufficiently large objects must behave classically. Any object larger than the critical length value based on the Hubble constant must behave classically except if Hubble flow suppression is present, and if that is present, then any object larger than a modified critical length based on a local effective Hubble constant must behave classically.

It is interesting that these different approaches to the question of a quantum-classical transition all arrive at rather similar expressions for the nominal threshold or critical size, but on the basis of very different reasoning. One derivation bases the transition on the presence of Hubble expansion velocities within the volume occupied by an extended object, whereas two other separate derivations do not require an extended object and ignore the explicit expansion of the universe but still do depend on a finite lifetime for the universe to derive comparable values for the critical size in conjunction with a quantum-classical transition.

We emphasize that all three separate approaches do arrive at essentially similar basic expressions for a threshold or critical size at which a quantum-classical transition could occur, which does appear to provide support for the basic validity of this result. The agreements between these three different approaches using conventional quantum mechanics and stochastic quantum theory are simply too powerful a set of clues to ignore. Together, they suggest that there is some validity to these conclusions and possibly also point to the existence of underlying connections between quantum mechanics and the behavior of the universe as a whole.

**C. L. Herzenberg**
e-mail: carol@herzenberg.net

q c tx wave packet dispersion.doc
9 June 2007 draft